**Designing high endurance $Hf_{0.5}Zr_{0.5}O_2$ capacitors through engineered recovery from fatigue for non-volatile ferroelectric memory and neuromorphic hardware**

Xinye Li, Padma Srivari and Sayani Majumdar *

Faculty of Information Technology and Communication Sciences, Tampere University, 33720 Tampere, Finland

E-mail: sayani.majumdar@tuni.fi

Heavy computational demands from artificial intelligence (AI) leads the research community to explore the design space for functional materials that can be used for high performance memory and neuromorphic computing hardware. Novel device technologies with specially engineered properties are under intense investigation to revolutionize information processing with brain-inspired computing primitives for ultra energy-efficient implementation of AI and machine learning tasks. Ferroelectric memories with ultra-low power and fast operation, non-volatile data retention and reliable switching to multiple polarization states promises one such option for non-volatile memory and synaptic weight elements in neuromorphic hardware. For quick adaptation of industry, new materials need complementary metal oxide semiconductor (CMOS) process compatibility which brings a whole new set of challenges and opportunities for advanced materials design. In this work, we report on designing of back-end-of-line compatible ferroelectric $Hf_{0.5}Zr_{0.5}O_2$ capacitors that are capable of recovery from fatigue multiple times reaching $2P_r > 40$ μC cm$^{-2}$ upon each retrieval. Our results indicate that with specifically engineered material stack and annealing protocols, it is possible to reach endurance exceeding $10^9$ cycles at room temperature that can lead to ultralow power ferroelectric non-volatile memory components or synaptic weight elements compatible with online training or inference tasks for neuromorphic computing.

**Key Points**: Recoverable fatigue, ferroelectric $Hf_{0.5}Zr_{0.5}O_2$ memories, control of ferroelectric crystallinity and interface, role of thermal engineering on leakage currents and endurance.

# 1. Introduction

New generation of non-volatile memories (NVM) capable of multibit operation are essential for densely integrated, beyond-CMOS computing hardware. [1, 2] Operation at ultra-low power and capability of working on biological principles of neuronal firing and synaptic plasticity of the brain, requires development of custom-engineered devices and their special programming protocols. The new devices utilize physics to realize energy-efficient artificial intelligence (AI) computation. Selection of the hardware building blocks rely on the materials and devices that can emulate the large dynamic conductance ranges for multibit operation and varied timescales of synaptic plasticity and neuronal signaling together with CMOS compatibility for their integration to standard semiconductor fabrication processes. From energy efficiency, non-volatile data retention and forming-free and fast switching perspective, ferroelectric (FE) memories have become an attractive choice for embedded memory and neuromorphic hardware. [3, 4]

Discovery of ferroelectricity, arising from the polar orthorhombic phase in CMOS compatible $Hf_{0.5}Zr_{0.5}O_2$ (HZO) thin films [5] has boosted this research field significantly. In recent times, three-terminal one-transistor one-capacitor (1T-1C) FE random-access memories (FeRAMs), [6] ferroelectric field-effect transistors (FeFETs) [7] or two-terminal ferroelectric tunnel junctions (FTJs) [8] have shown promising performances as non-volatile memory components with fast and reliable operation and ultralow power consumption. However, study on their potentials for scalability, nano-scale device performance and their 3D integration with CMOS readout and logic circuits are still at their early phases. [9] Moreover, in thin film polycrystalline FE devices, a significant leakage current exists that results in early fatigue and breakdown. [10] Low leakage current and high endurance is of utmost importance for their implementation in densely integrated non-volatile memory array and in neuromorphic hardware capable for on-line training.

Fluorite structure HZO has shown high remnant polarization ($P_r$) of 32 µC cm$^{-2}$ [11] in 10 nm films when the films were crystallized to polar orthorhombic phase using rapid thermal annealing (RTA) temperature of 700 °C. This RTA condition fails to meet the CMOS BEOL compatibility. Cheema et al. have shown that phase transition of fluorite structure HZO to the polar orthorhombic phase are energetically favorable at reduced thickness, [12] and some reports suggest that the crystallization can happen at a temperature range of 400-500 °C. [13] However, most of the reported results show that orthorhombic phase is more pronounced when RTA is above 500 °C. [14,15] Besides annealing temperatures, several fabrication parameters can affect the FE performance of thin film HZO such as oxygen pressure, [16] in-plane tensile stress [17] and material composition. [5,18] The current work focuses on improvement of endurance of metal-ferroelectric-insulator-metal (MFIM) capacitors at room temperature as well as programmable recovery from fatigue behavior by HZO microstructure and interface engineering. Major development is targeted on CMOS back-end-of-line (BEOL) compatibility, scalability, uniformity and high yield. The experimental results from prototypical thin film FE capacitors based on polycrystalline HZO show how the microstructure and interface engineering of the MFIM device can affect the FE polarization switching, leakage current, endurance fatigue and recovery characteristics, making it possible to design different functionalities needed for programmable memory and synaptic components for neuromorphic training and inference tasks. Since all memory and synaptic weight elements finally need integration with the CMOS logic circuits and one critical consideration for dense memory circuit is 3D vertical integration of memory and logic, we tested a range of temperatures for most suitable device properties finally focusing on nano fabrication processes that are compatible with monolithic CMOS BEOL integration. We find that devices with lower annealing temperatures of 450 $^0$C possess lower amount of orthorhombic phase, however, can still demonstrate a sizable 2$P_r$ of 37 µC cm$^{-2}$ at ±4.6V without any "wake-up" pre-pulsing. Additionally, these devices can endure >10$^9$ cycles of programming and erasing pulses and can recover from fatigue multiple times. Reduction of oxygen vacancy by adding another dielectric oxide layer in the stack can increase the 2$P_r$ to 66 µC cm$^{-2}$ at ±6V, however with a quicker set in of fatigue that is recoverable up to 90%.

   This work provides an important design guideline for improving the performance of ferroelectric device based non-volatile memory or synaptic weight elements with large number of electrically programmable and erasable states and long lifetime, sufficient for their utilization in deep neural network (DNN) accelerators or spiking neural networks (SNN) hardware.

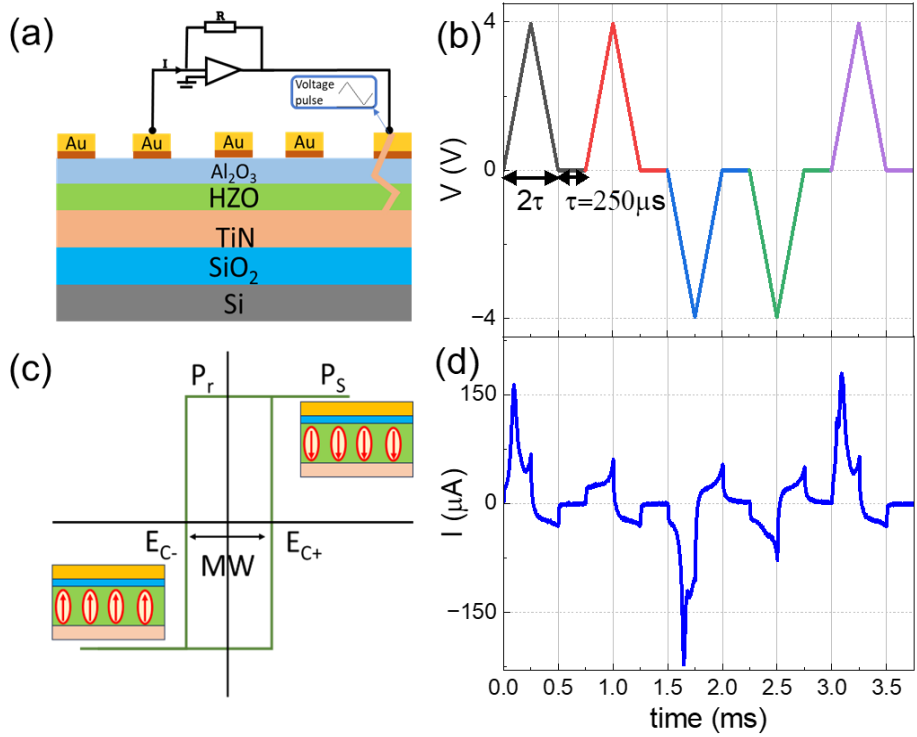

**Figure 1. (a)** Schematic diagram of the MFIM ferroelectric capacitor structure. Inset: triangular voltage sweep used in dynamic hysteresis measurements (DHM). **(b)** An example of PUND voltage pulsing at 4 V used for ferroelectric hysteresis measurements. **(c)** An ideal ferroelectric hysteresis with the key parameters for memory elements defined. **(d)** PUND current measured from S2 as an example.

## 2. Results and Discussion

### 2.1. Structural Characterization

**Figure 1** shows the MFIM capacitor structure used in the experiment together with their electrical characterization setup and a typical polarization hysteresis curve. Details of the fabrication and characterization methods are described in the Methods section, however, for introducing the nomenclature of the samples we briefly describe the sample fabrication conditions here. The MFIM capacitors were fabricated using Atomic Layer Deposition (ALD), with the bottom TiN (30 nm) electrode being grown by plasma-enhanced ALD (PEALD), followed by thermal ALD deposition of HZO (10 nm) and $Al_2O_3$ (1.2 nm) at 200 °C and rapid thermal annealing (RTA) of the samples for 30 seconds at annealing temperatures of 450 °C, 550 °C and 600 °C under nitrogen atmosphere (samples are named S1, S2 and S3 respectively). Finally, the FE capacitor structures was completed by evaporating Ti/Au as top electrode.

**Figure 2** shows grazing incidence x-ray diffraction (GIXRD) patterns of S1, S2 and S3 and approximate analytical estimation of different phases in the samples. The full range scan of 2θ from 10° to 80° shown in supplementary **Figure S1**. In **Figure 2 (a-c)**, we focus on the diffraction peak around 30° by doing a high-resolution scan around the peak. Previously, it has been reported that Zr doped $HfO_2$ can crystallize into monoclinic (*m*), orthorhombic (*o*), and tetragonal (*t*) phases [5] depending on annealing temperature [14], in-plane tensile stress from bottom electrode (BE), capping electrode material [17,19] and so on. [16] The formation of the *o*-phase leads to ferroelectricity while the *t*-phase favors antiferroelectricity in Zr-doped $HfO_2$. [5, 20] As shown in **Figure 2,** all our sample peaks match well with the HZO results reported in literature [21] where the Bragg peak positions and shapes suggest coexistence of monoclinic (*m*-phase, space group *P21/c*), tetragonal (*t*-phase, space group *P42/nmc*) and orthorhombic phases (*o*-phase, space group *Pca2₁*). It is important to point out here that due to the symmetric line shape of the narrow-range scan between 28° and 34°, it is difficult to clearly separate the *o* and *t* phase contributions by GIXRD alone. [12] However, a comparison of the phase ratio of the three samples, quantified by the same technique, can provide a fair estimation of the phase-fractions due to different thermal treatments. To quantify the contribution of each phase to the final diffraction peak around 30°, we fitted the curve by deconvolution of gaussian function with fixed peak at 30.4°, 30.8°, 31.6° that represent the *o*, *t* and *m* diffraction peak of HZO, respectively. [15] Calculation of the area under each curve shows that all the three samples have around 5% *m*-phase after RTA (**Figure 2 (d)**), confirming negligible percentage of non-polar phase fraction in all samples. S2, which was annealed at 550 °C, shows higher percentage of *o*-phase compared to S1 and S3, while S1 (RTA at 450 °C) shows higher percentage of *t*-phase compared to S2 and S3. All three samples are polycrystalline mixed phase with S2 being *o*-phase dominant, S1 being *t*-phase dominant while S3 shows almost similar amount of *o* and *t*-phases. Among all our samples, S2 has the smallest Bragg peak diffraction angle at 30.55° (**Figure 2(b)**) suggesting larger out-of-plane crystal size formed followed by S3 at 30.65°. This is expected considering the previous reports of higher orthorhombic phase formation due to higher annealing temperature. [14, 19] In our previous study, [21] we found that the preferential growth of TiN along *(200)_c* (cubic, space group $Fm\bar{3}m$) can positively influence the growth of the HZO *o*-phase. High amount of polar phase in S1, S2 and S3 also supports this conclusion.

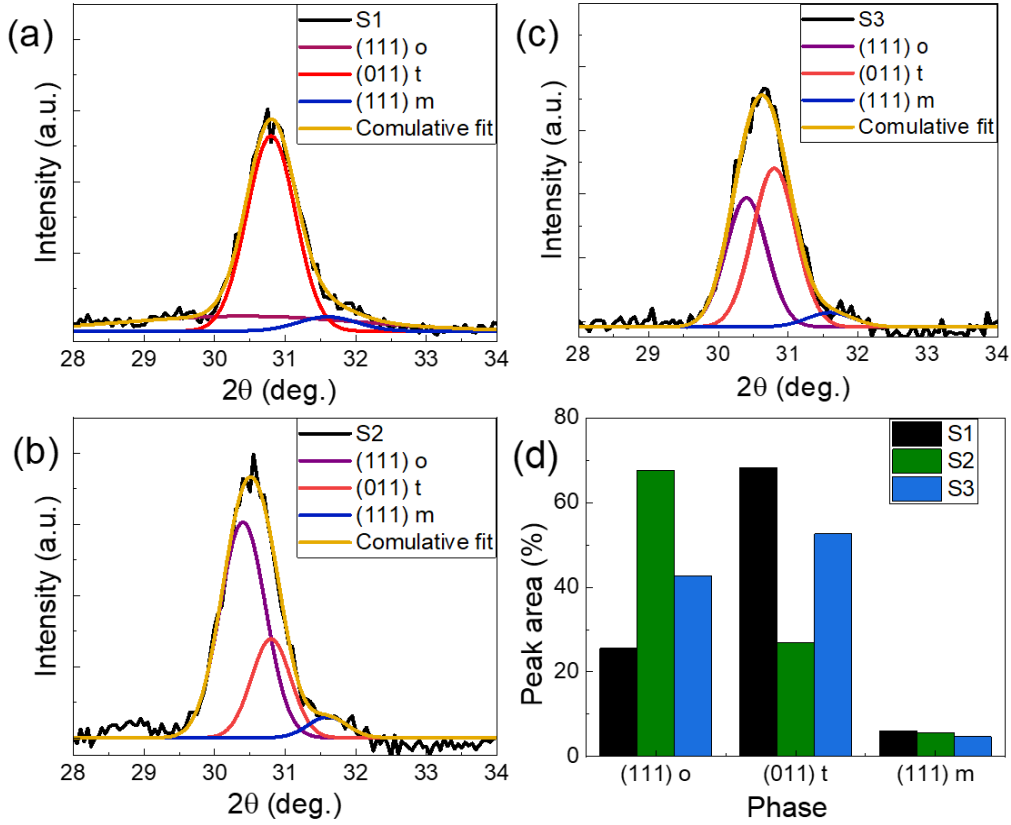

**Figure 2 (a), (b), (c)** GIXRD 2θ scans in range of 28° to 34° over the predicted *m* (-111), *o* (111) and *m* (111) peak positions of S1, S2 and, S3 respectively. Peak fitting of 2θ scans of S1, S2 and S3 with different reference diffraction peaks are used to calculate the ratio of different crystalline phases. **(d)** Comparison of *o*, *t* and *m*-phases in the three samples.

## 2.2. Electrical Characterization

### 2.2.1. Polarization (P) – Electric Field (E) Characteristics

**Figure 1(b)** shows the positive-up-negative-down (PUND) pulse parameters used for ferroelectric hysteresis measurements and **Figure 1(d)** shows the corresponding dynamic current response of the capacitors. In ferroelectric materials, under an external electric field, a net switchable polarization appears that is retained even after the electric field is removed, giving rise to non-volatile retention of the polarization, known as remnant polarization ($P_r$). With the increasing external electric field, the net polarization of the material increases due to more and more polarized domains aligning themselves in the direction of the external field. Beyond a critical field $E_C$, the coercive field of the material, most of the polarized domains align in the direction of the external field. Therefore, for $E>E_C$, the polarization starts to saturate at $P_S$ and remain at a certain value $P_r$ at zero bias. Similarly, when a negative bias is applied, polarization saturates at $-P_S$ with coercive field of $-E_C$ and remains at $-P_r$. This switchable polarization is used to store data in ferroelectric capacitors. Under ideal situation of fully screened polarization charges, the devices can maintain a square like hysteresis, as shown in **Figure 1(c)**. However, in practical devices with interfacial dead layer formation between metal-FE layer and other crystal defects, polarization loss happens resulting in rounded hysteresis. Besides, in ultra-thin film capacitors, needed for nano-scale electronic components a significant dielectric leakage current appears that affects the true estimation of the polarization charges and reduce device lifetime.

**Figure 3 (a), (b), (c)** show the dynamic current-voltage (*I-V*) loops and **(d), (e), (f)** are polarization-electric field (*P-E*) hysteresis of S1, S2 and S3 measured by the PUND technique. All data are from the pristine HZO capacitors without any wake-up cycling. For PUND measurements, we used triangular pulses of 1 kHz frequency and a voltage amplitude of up to ±4.6 V. Each PUND measurement sequence consists of one pre-poling pulse, read pulse and subsequently a rewrite pulse. In PUND, the first pulse of each pair contains all current contributions, and second pulse of the pair arises from the dielectric and leakage contribution of the device. [22] The final *P-E* hysteresis is calculated by subtracting the current response from the second up pulse from the last positive voltage pulse since the first positive pulse has memory from the pre-poling, and the fourth down pulse from the third negative pulse. The polarization charges measured using PUND, thus, does not consider the dielectric and leakage current contributions and gives more accurate values of $P_S$ and $P_r$.

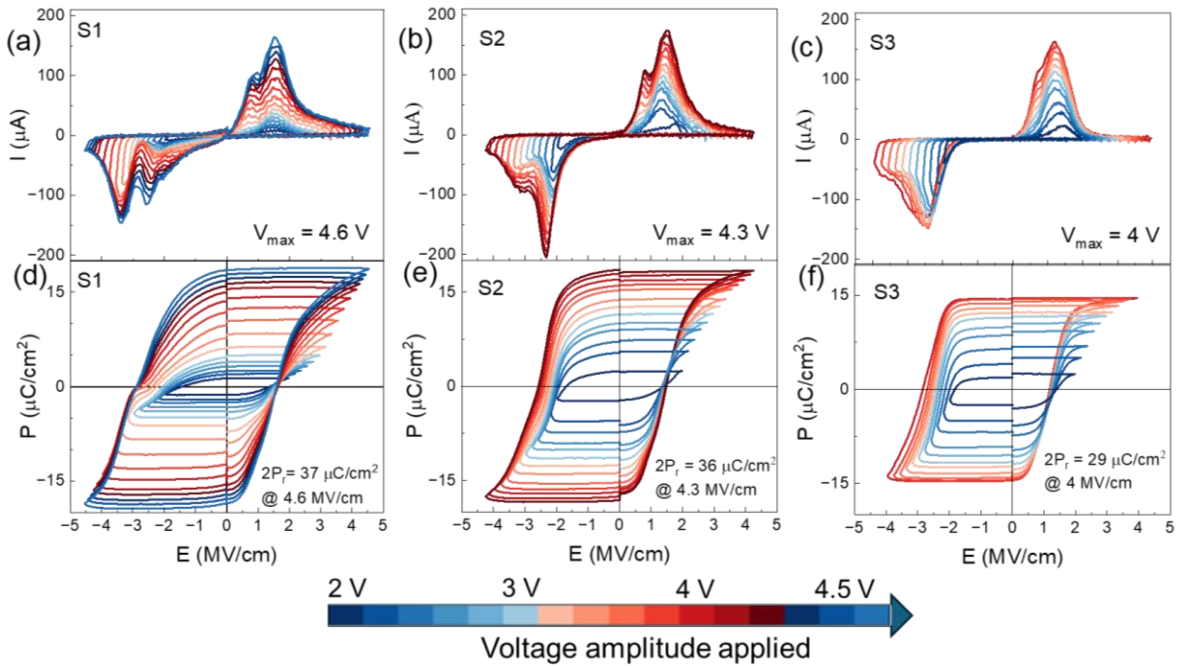

**Figure 3.** *I-V* **(a), (b), (c)**, and *P-E* **(d), (e), (f)** plots for samples S1, S2 and S3 respectively. Measurements were performed with PUND pulsing scheme with triangular pulses of 1 kHz frequency.

The S1, S2 and S3 show distinct differences in the displacement current features near the $E_C$. While S1 shows a broad transition with double peaks as shown in **Figure 3 (a)**, the transition for S2 and S3 is much sharper. This double peak feature becomes visible in the $P – E$ loop as well where S1 shows a slightly distorted $P – E$ hysteresis while that for S2 and S3 are rather sharp FE hysteresis. This separated switching current peaks might have different origins: 1) coexistence of tetragonal and orthorhombic phases in HZO with the tetragonal phase leading to an antiferroelectric like double-peak switching [20] or 2) structural defect-rich sample where rapid domain rotation gets restricted due to presence of large number of domain pinning centers leading to multiple $E_C$s that becomes visible through multiple displacement current peaks or coexistence of both. To verify the dominant process in our samples, we did field cycling of the samples. The double peak feature of S1 and S2 disappears after $10^7$ and $10^6$ cycles of pulsing, respectively with ±3V square pulses of 100kHz frequency, indicating field-induced crystallization from $t$ to $o$-phase of HZO happens in the samples that are annealed at lower temperature, as also previously observed by Lederer et al. [20] However, even the 450 $^0$C annealed samples don't show a pinched hysteresis in the pristine phase when measured with dynamic hysteresis measurements (DHM), suggesting $t$-phase not to be a dominating factor. The second factor, *i.e.,* high density of structural defects, is a major factor in S1 and partially in S2 which brings both advantages and challenges and have been discussed in later sections of the article.

For all three samples, the *P-E* loops were measured varying the pulse amplitude from ±2 V to ±4.6 V showing the possibility of multiple stable polarization states. Increasing voltage amplitude manifested in an increasing switching current, leading to higher saturation and remnant polarization. Almost linear increase in $P_r$ value in the range of 2.5 to 18 µC cm$^{-2}$ was observed for S1 and S2 due to increasing pulse amplitude while for S3 a steeper rise in $P_r$ value is obtained in the lower field regime followed by a saturating trend. However, the $P_r$ is limited to 15 µC cm$^{-2}$ for S3 since hard breakdown happens when the applied voltage is above 4 V, whereas S1 and S2 can endure until 4.6 V and 4.3 V, respectively, as shown in **Figure 3 (a,b)**. Such phenomenon is in line with the previous reports that with increasing annealing temperature there is an increasing leakage contribution to the MFM device, [19, 23, 24] which becomes dominant current flow at a higher electric field and short the device due to hard breakdown of the HZO layer. S1 and S2 shows the highest $P_r$ of ~18 µC cm$^{-2}$ at 4.6 V and 4.3 V respectively. This clearly indicates that due to RTA at 550 °C, higher fraction of HZO gets crystallized to the ferroelectric orthorhombic phase compared to S1 and S3, resulting in higher $P_r$ and lower $E_C$. Although S1, which was annealed at only 450 °C, is not fully crystallized to

*o*-phase, application of higher voltage can lead to comparable $P_r$ value since S1 can endure higher bias stress due to lower crystallinity and oxygen vacancies.

The $E_C$ values for positive and negative voltage side of the *P-E* loops are different in all samples showing the imprint effect of the ferroelectric capacitors (**Figure 3**). In FE capacitors, imprint effect arises from dissimilar electrode configuration [25] on either side of the FE, stress induced from electrode lattice mismatch [17] or from asymmetric charge trapping at the ferroelectric electrode interface by defects. [14] In the present experiments, a dielectric (DE) layer of $Al_2O_3$ is intentionally added in all the MFM devices. The choice of $Al_2O_3$ capping layer was motivated by the fact that this ultrathin oxide ensures good ferroelectricity in HZO devices [21] that were not annealed in presence of TiN top electrode, which was considered mandatory in earlier works. The choice of 1 nm thickness of $Al_2O_3$ layer was motivated as a compromise between its contribution to depolarization field consequently causing retention loss, [22] and serving as a tunnel barrier to ensure the stable ferroelectric switching. [26, 18] The 1 nm dielectric layer at the top interface inhibits the down to up polarization rotation leading to higher value of $-E_C$ compared to $+E_C$. Additionally, the change in $E_C$ due to different applied pulse amplitude is more pronounced on the negative side suggesting more restricted domain rotation from down to up direction resulting from trapped charge related domain pinning sites at the HZO-$Al_2O_3$ interface. Due to same composition of the capacitor stack, the imprint effect is supposed to show identical values in all samples. In reality, however, crystal structures, grain orientations and charged oxygen vacancies formed during RTA at different temperatures are different and subsequently pulsing experiments contributing to formation of charge trapping and domain pinning sites also vary, leading to different performance of S1, S2, S3 with S2 showing the least imprint and sharpest polarization switching.

*2.2.2. Wafer-scale Uniformity*

For implementation of the devices in larger circuits and for cost-effective industry-scale production, it is important to have information of wafer-scale uniformity of the device performance. It is previously found that fabrication process parameters affect size of crystallites, thereby affecting the wafer-scale uniformity of the devices. [27] In order to show the effect of RTA temperatures on the wafer-scale uniformity, in **Figure 4** we plot the distribution of the $P_r$ values for multiple devices from S1, S2 and S3 measured from different locations all over one-quarter of a wafer. Based on the position of the sample on wafer, the $P_r$ values can range from 2 to 15 µC cm$^{-2}$ when measured with ±3V DHM pulses of 1 kHz frequency, represented by the color scale. The devices shorted during fabrication and measurement processes have been represented in black color. As a general observation, the edges of the wafers show higher density of low-performance or shorted devices, likely caused by mechanical defects caused by wafer handling, non-uniformity of growth of the PEALD grown TiN bottom electrode and non-uniform thermal gradient during RTA process. The hypothesis of a thermal gradient during RTA is derived from the fact that in all 3 samples, the center of the quarter showed most uniform device performance. From the color map, it becomes clear that S1 and S2 show less wafer-scale uniformity with the $P_r$ values ranging from 2.4 to 8.7 µC cm$^{-2}$ for S1 and 3 to 15.2 µC cm$^{-2}$ for S2 respectively. Highest number of devices shows $P_r$ of 6 µC cm$^{-2}$ for S1 and 8 µC cm$^{-2}$ for S2. The higher temperature annealed sample S3 is comparatively more uniform with $P_r$ values ranging from 5.9 to 12.9 µC cm$^{-2}$ and a peak around 11 µC cm$^{-2}$. There exists certain degree of variation in all 3 samples. The supplementary **Figure S2** shows DHM from multiple devices from S1, S2 and S3 for a quick comparison. Contour plots in **Figure 4** (**a-c**) confirm that lower temperature annealing contributes to low $P_r$ all over the wafer and narrower distribution of $P_r$ values throughout the wafer. Higher temperature annealing, on the other hand, leads to a larger $P_r$ value with higher variation of polarization values throughout the sample. Due to high leakage currents, more shorted junctions are found in S3. However, within 1x1 or 2x2 cm² proximity in the middle of the wafer, it's possible to find a narrower distribution of device performance matrices in S3.

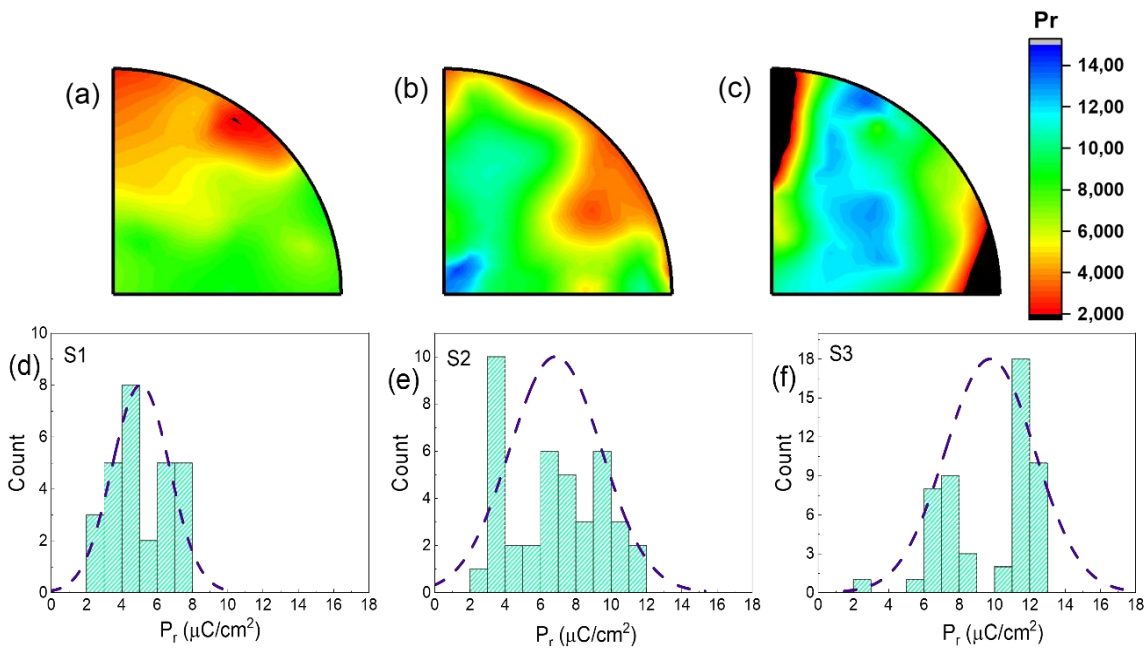

**Figure 4.** Mapping of the remnant polarization ($P_r$) values from the measured devices from one-quarter of a 150 mm wafer. All $P_r$ values are measured using dynamic hysteresis measurement (DHM) with ±3V, 1 kHz voltage applied. **(a), (b), (c)** are contour plot showing physical location of the high $P_r$ and low $P_r$ devices while **(d), (e), (f)** shows statistical distribution of $P_r$ data from samples S1, S2, S3 respectively.

This trend of higher temperature annealed samples having higher $P_r$ values agrees well with the GIXRD data and previous*ly* reported trend [19, 14, 11] that with increasing RTA temperature (until 800 °C [24]), increased amount of *o*-phase crystallized can lead to increment of $P_r$ values. However, from the uniformity point of view, S1 enjoys the low standard deviation value 1.7, which is lower than 2.8 for S2 and 2.5 for S3. S3, on the other hand, shows highest number of high $P_r$ devices around the center part of the sample, which leads to the possibility of densely packed high-performance circuit fabrication around this region. Low $P_r$ samples around the edge of the wafers suggest incomplete crystallization around these areas that would require further investigation and process parameter optimization before a higher uniformity is reached.

*2.2.3. Endurance Characteristics*

For neuromorphic computing tasks, such as online training, one important device parameter is very high endurance of the synaptic devices since devices need to switch for more than $10^9$ cycles to learn the correct synaptic weight. [28] To estimate the endurance properties of our devices, the fatigue measurement tests on the 3 samples with the following protocol were carried out.

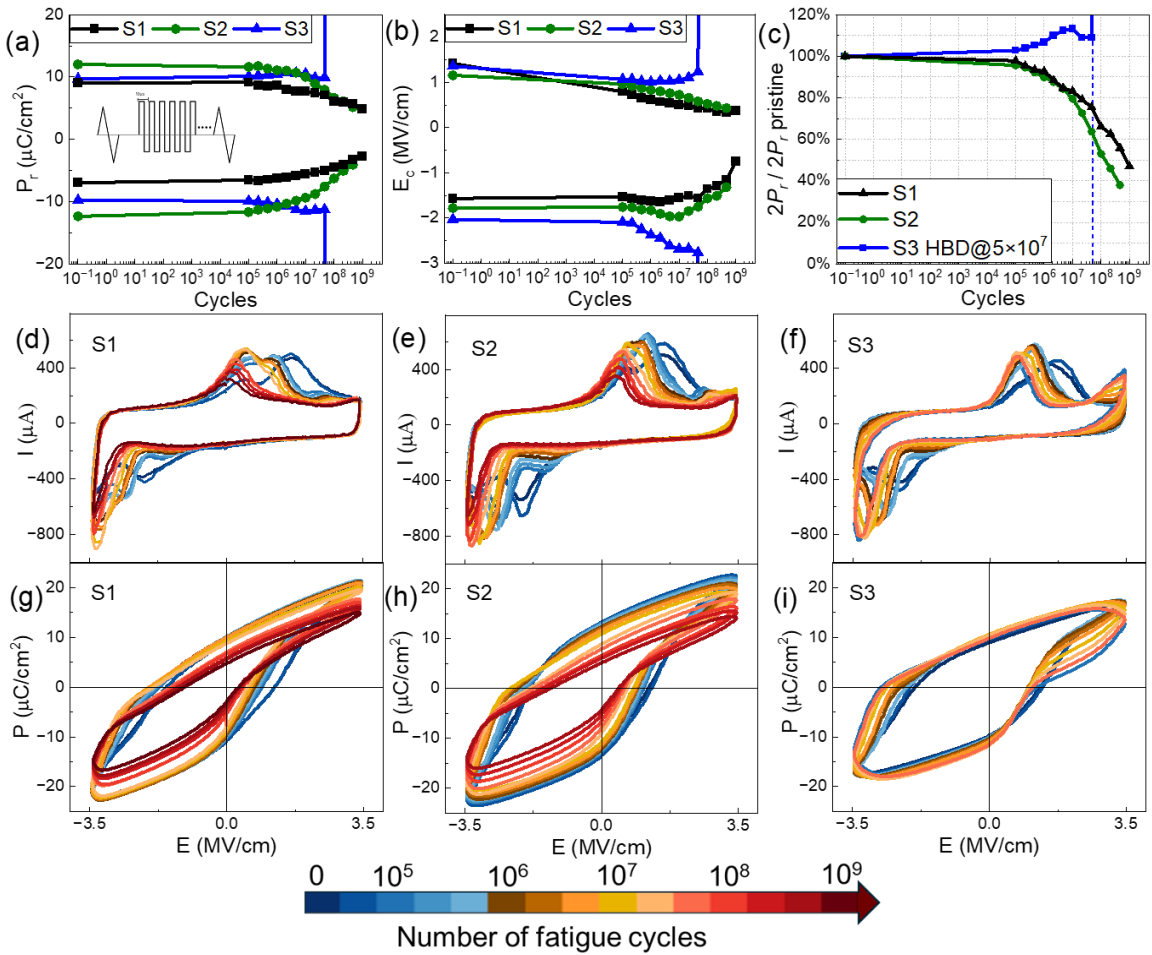

**Figure 5.** Endurance tests performed on S1, S2 and S3 junctions showing evolution of **(a)** $P_r$ ($P_{r+}$ and $P_{r-}$) and **(b)** coercive fields ($E_{C+}$ and $E_{C-}$) as a function of cycle number. **(c)** Normalized $2P_r$ values of S1, S2 and S3 in accordance with fatigue cycles number. Inset of (a) shows rectangular 3V, 100 kHz voltage pulsing scheme used for all fatigue tests. DHM measurements were performed (at each data point) with triangular pulse of 3.5V magnitude at 1kHz frequency. **(d-i)** Evolution of the DHM: *I-V* hysteresis (**d-f**), *P-E* hysteresis (**g-i**) shape from pristine to fatigued samples, showing a reduced $P_r$ and narrowing memory window (MW) in S1 and S2 while for S3 only visible degradation mechanism is imprint.

As shown in **Figure 5(a)** inset image, devices were pulsed with rectangular pulses of ±3 V, 100 kHz frequency. In order to investigate the evolution of FE hysteresis over several field cycling, every 1/3$^{rd}$ of a decade, one DHM measurement was performed with triangular pulses of ±3.5 V amplitude and 1 kHz frequency. S1 and S2 devices were able to endure long pulsing cycles exceeding $10^9$ cycles (when DHM was measured at 3V) with diminishing $P_r$ and $E_C$ values. However, for S3, devices showed hard breakdown (HBD) after $4*10^7$ cycles as shown in **Figure 5 (a-c)** with a trend of increasing $P_{r-}$ and $E_C$ due to field cycling. A gradual fatigue behavior started setting in after almost $10^7$ cycles in S1. In comparison, S2 started to show faster development of fatigue (after $10^6$ cycles) under the same pulsing protocol. The $2P_r$ value (sum of $P_{r+}$ and $P_{r-}$) decayed to 63% of the pristine $2P_r$ at $2*10^8$ and $4*10^7$ number of cycles for S1 and S2 respectively in **Figure 5 (c)**. At the end of the fatigue test for S2 ($4*10^8$ cycles), $2P_r$ further decayed to 37%, in contrast, S1 remain 55% of pristine $2P_r$ after the same number of cycles and was able to sustain further pulsing cycles exceeding $10^9$.

**Figure 5 (d-i)** shows the dynamic *I-V* and *P-E* hysteresis before and during the fatigue tests. The dynamic *I-V* hysteresis during the fatigue tests (**Figure 5 (d-f)**) show that with increasing number of field cycling, the leakage current contribution starts to increase, which is shown in **Figure 5 (h-i)** as a shifting (narrowing for S1 and S2 is discussed later) of the *P-E* hysteresis towards negative polarization direction caused by movement of oxygen vacancies under continuous bias stress. In all three cases, leakage current starts to increase with increasing field cycling, mainly on the negative voltage side, confirming the role of positively charged oxygen vacancy migration under repeated field cycling and eventual conducting path formation to be one major contributing factor for leakage current in thin film HZO capacitors. It results in degradation of $2P_r$ and $E_C$s in S1 and S2 while causing HBD in S3 samples. HBD in S3 is most likely the result of filamentary path formation of oxygen vacancies that are strong enough not to get ruptured under opposite bias causing a permanent conduction channel between the top and bottom electrodes.

It is observed in the evolution of DHMs of S1 and S2 in **Figure 5 (d, e, f),** that the double peak feature merges into one switching current peak while shifting to the smaller voltage value followed by a degradation of switching current on both positive and negative field side, whereas S3 shows HBD before showing any significant degradation. The degradation is shown as a constantly decreasing $P_{r+}$, $E_{C+}$ and a first increasing then decreasing $E_{C-}$ in S1 and S2 **Figure 5 (g, h)**. Such shift of switching current on the voltage axis and degradation of $P_r$ and MW (sum of $E_{C+}$ and $E_{C-}$) upon continuous pulsing can be attributed to various effects that come into play during continuous bias stressing: 1) mixed-phase polycrystalline HZO (in this case S1 and S2) undergoing field-induced crystallization to *o*-phase, so that the external electric field needed to switch polarization is less than that of a pristine sample 2) the reduced $P_r$ and imprint phenomenon can be attributed to the asymmetric charge trapping by defects[14] at the FE-DE interface in this study, HZO-TiO$_x$N$_y$ (TiON) and HZO-Al$_2$O$_3$. The latter factor is discussed in section 2.2.1 and it is similar in all S1, S2, S3 samples. Different endurance performance and fatigue mechanism can be therefore attributed mainly to the former factor HZO-TiON interface and different amounts oxygen vacancy formation due to different oxidation of TiN caused by different RTA temperatures.

In case of S3, only visible fatigue mechanism is imprint, which is shown as continuously increasing $E_{C+}$ and $E_{C-}$ in **Figure 5 (i)** and no degradation of $2P_r$ and MW narrowing happens until the breakdown, as also can be seen from **Figure 5(a-c)**. Moreover, a hard breakdown appeared before development of severe fatigue (degradation of $P_{r+}$). This is in line with the data in **Figure 3 (c, f)** where S3 was unable to handle larger PUND pulse amplitudes, indicating larger leakage current in high temperature annealed samples. With increasing pulse amplitude in fatigue measurements, devices started showing increased fatigue, and could not sustain high number of field cycling. For S3, the *P*-E loop starts shifting on the voltage axis due to repeated field cycling showing increased imprint effect arising from charged defect generation during field cycling.

*2.2.4. Leakage Current – Effect of Microstructure and Interface*
The *I-V* and *P-E* data in **Figure 3** are the results from PUND measurements where the leakage current components are subtracted from the total current to correctly estimate the displacement current component. In real applications, however, the leakage component from the ultra-thin capacitors cannot be subtracted and therefore is a critical parameter that needs to be considered while evaluating device performance. As seen from the GIXRD data (**Figure 2**) and the dynamic *I-V* measurements (**Figure 3**), with increasing annealing temperature, the ferroelectric orthorhombic phase increases. From the static current – voltage measurements (**Figure S3**), we find that static current of the sample S2 is almost an order of magnitude higher compared to S1 at 2V bias and almost 2 orders of magnitude higher for smaller bias. As discussed previously, it has been found previously that low temperature annealing leads to a semi-crystalline phase [20] in HZO with formation of less oxygen vacancy. [11] Presence of the amorphous phase leads to reduced leakage current paths and reduced oxygen vacancy formation causes less trap assisted tunnelling. Both effects contribute to decreased leakage current density in sample S1. Higher crystallinity of HZO leads to more randomly oriented grains and grain boundaries that results in higher amount of leakage current paths through the device resulting in quicker fatigue and breakdown compared to devices with lower polarization value with more amorphous oxide phase. S3 shows almost similar leakage current at higher voltage as S2, however a lower leakage at low bias range.

To investigate the role of the TiN-FE interface in modifying the leakage currents, we prepared another sample S4 based on the same fabrication procedure of S1, however, in S4, TiN was intentionally placed in the air for 7 days before HZO growth. This intentional oxidation of TiN into TiON gave us the opportunity to investigate the effect of having an oxygen reservoir layer at the HZO- bottom electrode interface that could modify the vacancy formation affecting the leakage currents in return. **Figure 6** shows the *I-V* and *P-V* characteristics of sample S4 showing a higher switching voltage on the positive voltage side (**Figure 6**(a)) and much higher $P_r$ of 33 $\mu C\ cm^{-2}$ (**Figure 6**(b)) in the sample due to the ability of S4 to withstand bias stress of ±6V. In comparison to S1, the devices showed identical $-E_C$ but increased $+E_C$, suggesting a significant voltage drop (of ~2V) across the bottom TiON layer. Most importantly, the leakage current in S4 decreased significantly compared to S1 and S3 in quasi-static range (**Figure 6(d)**). The static *I-V* data in **Figure S3** asserted this further by showing almost 3 orders of magnitude less leakage current in S4 in comparison to S2 and S3 and more than 2 orders of magnitude for S1 at 3V. This reduced leakage current can be attributed to the formation of less oxygen vacancies during the RTA step in presence of the oxygen reservoir layer at the bottom interface and an additional non-polar barrier formation preventing trap-assisted tunneling. This reduced leakage current resulted in long device lifetime, however, a large fatigue started to settle in after $10^5$ cycles of operation (**Figure 6(e)**). This fatigue, however, is found recoverable with a few DHM cycling with a slightly higher voltage amplitude compared to the pulsing voltage amplitude. The mechanism of fatigue in various kind of thin film ferroelectric devices and recovery from fatigue throws light on the interplay of charge trapping at the FE-DE interface and role of oxygen vacancies and can act as an important guideline in designing next generation of non-volatile memories. In the next section, we discuss the phenomenon in detail.

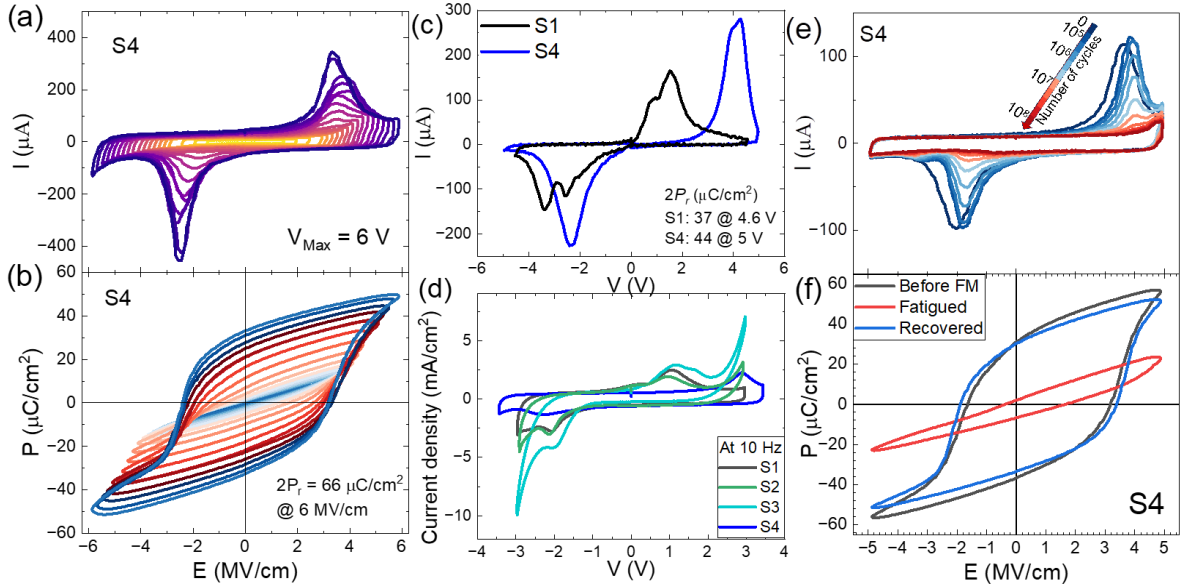

**Figure 6.** FE characteristics of S4: *I-V* **(a)** and *P-E* hysteresis **(b)** measured with DHM at 1 kHz with various voltage amplitudes showing the devices have a larger memory window and can withstand large electric fields of 6 MV/cm giving a high 2Pr value of 66 μC cm$^{-2}$. **(c)** Comparison between S1 and S4: *I-V* plots measured with PUND showing shift in positive voltage peak due to the presence of oxide layer at the bottom electrode, **(d)** low-frequency leakage currents with samples S1 and S3, showing leakage is significantly reduced in S4 **(e)** *I-V* plots from monitor DHMs (5 V, 1 kHz) during the fatigue measurement up to $10^8$ cycles, **(f)** DHMs measured before pulsing (black lines), after $10^7$ cycles of pulsing (red lines), after recovery (blue lines) with high voltage amplitude DHMs.

*2.2.5. Recovery from fatigue: Charge trapping and de-trapping at the HZO – dielectric interface*

**Figure 7** shows the fatigue of $P_r$ over pulsing cycles and their recovery trends in S1, S2 and S4 devices. It is found for all devices, a recovery from the fatigued $P_r$ values is possible by applying higher external electric field sweep with a custom protocol. For the recovery characterization, fatigue measurement (FM) was carried out the same way as shown in *2.2.3* with intermediate gaps while fatigued states set in and applying several DHM pulsing cycles at certain voltage amplitude for recovery. The $2P_r$ of S1 as shown in **Figure 7(a)** decayed to 11.8 μC cm$^{-2}$ after the first $10^8$ fatigue pulsing cycles. The recovery is carried out by applying dynamic hysteresis measurements up to 10 times, which consist of triangle voltage sweep at 1 kHz with a higher voltage amplitude (in case of S1 ±3.6 V, ±3.7 V, ±3.8V) compared to the pulsing voltage amplitude. After 3 DHM cycles, recovery from fatigue state was obtained with $P_{r+}$ of S1 fully recovered, whereas $P_{r-}$ is only 70% recovered, as shown in **Figure 7(a)**. The recovery of S2 is performed in the same way using DHMs with ±3.7 V, ±3.8 V, ±4V after $10^7$ pulsing cycles. Similar to S1, after 3 DHM cycles recovery from fatigue state was obtained with $P_{r+}$ of S2 recovering 96% compared to 79% of $P_{r-}$, as shown in **Figure 7(b)**. The loss of S1 and S2 from fatigue test, which cannot be recovered is the result of the charge carriers trapped in HZO-Al$_2$O$_3$ interfaces those are unable to respond to the applied recovery pulse sequences and prevent a full or higher percentage of recovery (**Figure S4 (a,b)**).

Since the MW for S4 is larger (as shown in **Figure 6**), voltage amplitude of ±5 V was applied for monitoring fatigue and recovery of S4. A recovery DHMs with ±5.5 V, ±5.7 V was applied after 10$^7$ voltage pulses in S4. As shown in **Figure 7**(c), the devices show a quicker set in of fatigue meanwhile a better recovery of 93% of pristine values on both up and downside of remnant polarization. As discussed in *2.2.3*, S2 decays relatively faster than S1, which is in line with the data in **Figure 7 (e)**. Nevertheless, after the recovery, the fatigue of all samples seems to have their pristine states nearly back and many longer pulsing cycles can be achieved.

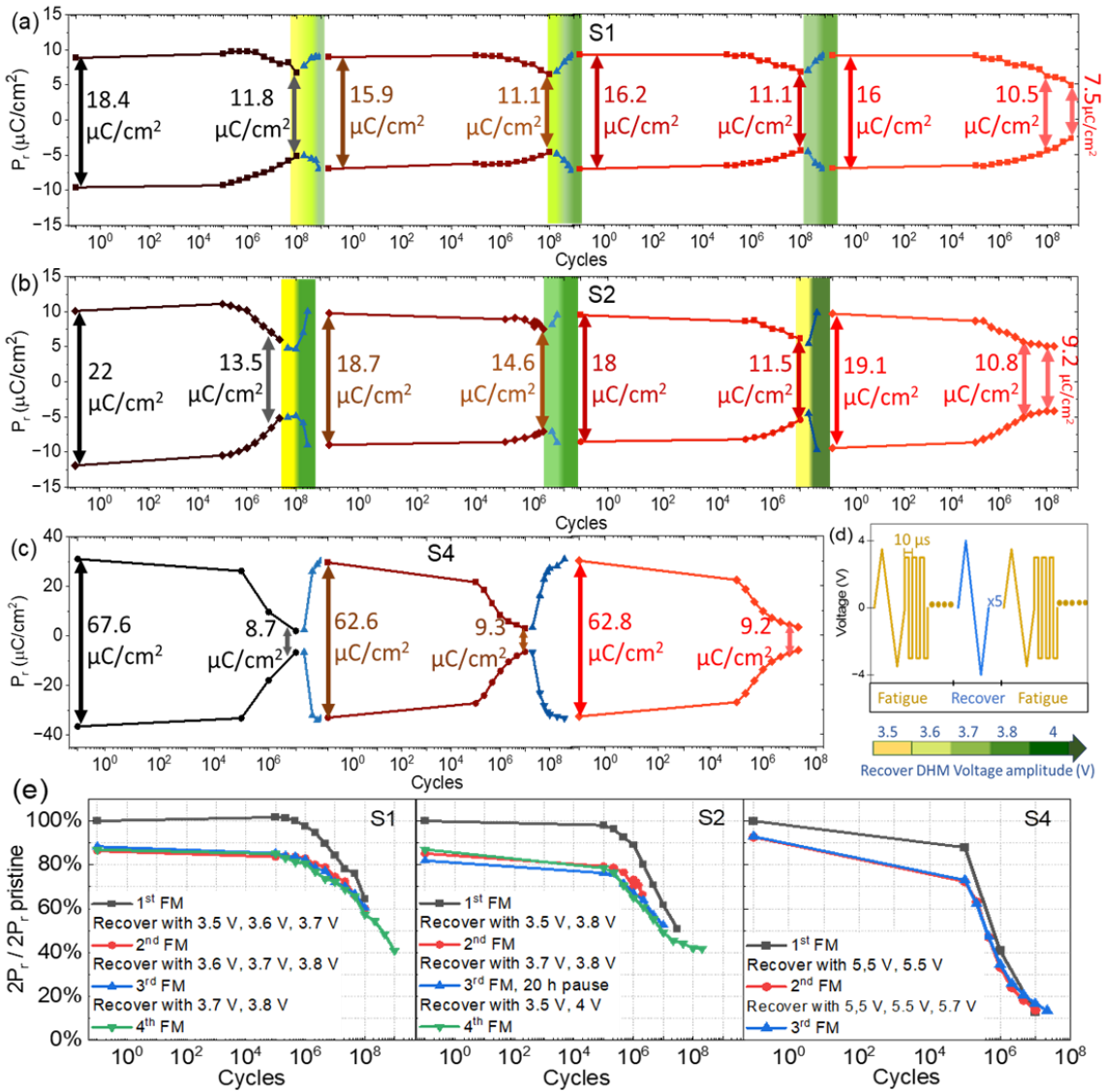

**Figure 7.** Recovery from fatigue measurements performed on S1 **(a)**, S2 **(b)**, and S4 **(c)** junctions showing evolution of $P_r$ as a function of cycle number. Monitoring DHM was carried

out at 3.5 V, 1 kHz for S1 and S2 and at 5 V, 1 kHz for S4 at each data point. (**d**) Pulsing schemes employed for fatigue and recovery measurements. Magnitude of the recovery pulses are shown by the color code. (**e**) Normalized $2P_r$ values of S1, S2 and S4 as function of fatigue cycles numbers after 1st, 2nd, 3rd recoveries showing same $P_r$ states can be achieved reproducibly after each fatigue and recovery cycle.

Recovery from fatigue in HZO devices have been reported previously. [29, 30, 31, 32] However, a microstructure and interface dependence of the effect and physical interpretation of the process is still missing. Here, we analyze the fatigue and recovery in our samples considering different electrostatic contributions and provide a guideline for the improvement of fatigue and recovery process through material and stack engineering.

The effective electric field across the ferroelectric layer in an MFIM capacitor, $E_F$ is given by,

$$E_F = \frac{V}{t_F}\left[1 + \frac{C_F}{C_{int}}\right]^{-1} - \frac{P+\sigma}{\varepsilon_0 \varepsilon_F}\left[1 + \frac{C_{int}}{C_F}\right]^{-1} \qquad (1)$$

where $P$ is the polarization order parameter, $V$ is the applied bias on the MFIM, and $C_F$, $C_{int}$ are the capacitances of the ferroelectric and interfacial layers, respectively and $t_F$ is the FE thickness. The first term in the right-hand side of the equation accounts for the reduction of the volage in the ferroelectric by a factor $[1+C_F/C_{int}]^{-1}$, as part of the voltage drops on interfacial insulator layers. The second term in the right-hand side of Eq. (1) is the electric field built due to uncompensated charges in the device that becomes significant during field cycling. The surface charge σ is the charge trapped at pre-existing defects at the interface. During field cycling, charge injection and trapping occur inside the FE layer and at the HZO/Al$_2$O$_3$ and HZO/TiON layers increasing the σ, reducing the effective field $E_F$ significantly. A low dielectric constant interface oxide reduces the $C_{int}$ effectively reducing the field across the FE. This reduced field switches only small number of polarized domains, leading to smaller $P_r$ switching, but gives the devices an exceedingly long lifetime. The smaller $P_r$ gets manifested in fatigue. Once the trapped electrons are de-trapped using larger magnitude pulses, σ is reduced and the devices show their original $P_r$ and show similar field cycling effects. For sample S3, the second term in the equation is not significant due to lower number of crystalline defects.

For a clearer understanding of the effect of interface trap densities on the observed fatigue and recovery behavior, we measured capacitance as a function of frequency (*C-f*) at zero dc bias for S1, S2 and S4, both in their pristine and fatigued states (**Figure S5**). Interface trap density ($D_{it}$) is generally calculated from the frequency dispersion of the capacitance. For our samples, S1 and S2 showed no change in zero bias capacitance values at the frequency range of 1-10 kHz while for S4, in the measured temperature range, the capacitance values changed slightly from the pristine to the fatigued sample showing different nature of charge trapping in the fatigue process for samples without or with double DE barrier. Both S1 and S2 shows a hysteresis in the *C-f* data for increasing and decreasing *f* scan direction in the pristine sample while the hysteresis disappeared in the fatigued samples. For S4, there were no observed hysteresis in the C-*f* scan, neither for the pristine nor the fatigued sample. This trend, together with the fatigue and recovery data, throws light on the underlying mechanism for the nature of charge trapping in the samples with single or double DE oxide interface. For S1 and S2, disappearance of C-*f* hysteresis in the fatigued sample shows frequency dispersion of capacitance is polarization dependent while in S4, polarization state does not play a significant role in frequency dispersion of capacitance. From the analysis of the leakage, capacitance and fatigue and recovery data, we then come to a unified picture that is schematically shown in **Figure 8**. In S1, S2 and S3, oxygen vacancies are formed during the RTA step as the TiN scavenges oxygen from HZO layer leading to a high density of vacancy related defects in the HZO layer. In comparison, much less vacancies are formed in S4 due to the thicker oxide layer at the bottom interface supplying oxygen to the TiN. During field cycling, S1 gets lot of charged vacancies distributed throughout the HZO layer that causes charge trapping centers and pinning FE domains at these traps. In S4, interface defect related traps exceed the vacancy related traps due to lattice mismatch of FE and DE layers that causes the domain pinning. Also double DE interface reduces the effective voltage drop across the FE. Interface traps, reflected by the *C-f* measurements, causes quicker fatigue but higher recovery. More in-depth characterization and modelling of charge trapping and de-trapping in these samples are currently underway.

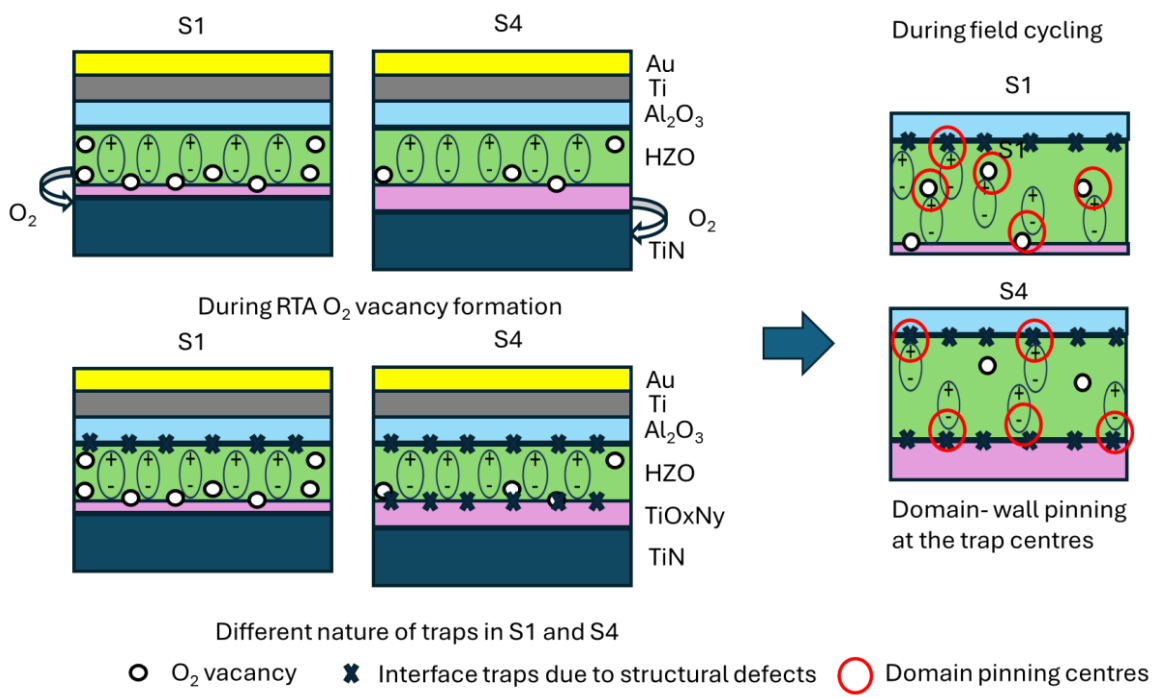

**Figure 8.** Schematic diagram of different defect states and charge trapping mechanism in S1 and S4 samples during pristine and fatigued states.

## 3. Application in Memory and Neuromorphic hardware

This result has significant implication for fabrication of densely packed 1T-1C FeRAM, 1T FeFET or capacitive memory arrays. In the semiconductor industry, dynamic random-access memories (DRAM) or static random-access memories (SRAM) are most used memories as volatile memories while NAND flash is most used as non-volatile memories. DRAMs operate with lower voltage showing fast read/write operations, however, with fast data loss that requires frequent refreshing of the memory. Flash memories operate at a higher voltage that restricts the device lifetime to $10^5$ cycles of operation. FE capacitor based FeRAMs can combine best of both worlds by performing fast read/write operations, non-volatile data retention and long device lifetime. Our current observation provides a design clue for high-performance non-volatile memory and neuromorphic circuits. While high $P_r$ values, lower switching voltage, fast switching times are the more desirable parameters for digital non-volatile memory components, for neuromorphic circuits requirements are more analog states, reproducibility of switching to multi-level conductance and endurance exceeding $10^9$ cycles for online training while for inference tasks a long data retention time is essential. Our results suggest, by low temperature crystallization of the ferroelectric layer, it is possible to design BEoL integrated 1T-1C FeRAMs with multibit operation, high endurance and recoverable fatigue that makes them suitable for online training works in neuromorphic circuits.

In our previous studies with single crystalline perovskite oxide $BaTiO_3$ and $Pb_{1-x}Zr_xTiO_3$ (PZT) based FTJs, the vital role of oxygen vacancy migration in ultrathin FE films was investigated that gave rise to large resistive switching. [33, 34] For polycrystalline polymer ferroelectric FTJs, it was found that through modification of microcrystalline structure of the ferroelectric, it is possible to have ultrafast switching (<20 ns) and controllable domain dynamics [35] that can result in efficient online training for a multilayer perception based neural network. [36, 37] Besides non-volatile memory functions, proper engineering of the depolarizing field at the ferroelectric-electrode interface can lead to control of the polarization relaxation in such a way that a programmable synaptic plasticity time constant [38] or leaky integrate-and-fire (LIF) functions can be achieved [39] using the same ferroelectric device that can significantly reduce fabrication process complexity and cost of the fully-connected neural network. For HZO, multibit operation with varying voltage amplitude, shown in this work and in ref. [21] leads the way for designing analog FE memory array. A SPICE compatible Jiles-Atherton model has been developed [40] for accurately predicting the circuit level performance of the HZO devices.

A crossbar array of ferroelectric capacitors, FTJs or FeFETs can serve as analog compute core, performing the vector matrix multiplication (VMM) operation by utilizing the Ohm's law and Kirchoff's current law leading to efficient parallel weight updates that significantly improves energy efficiency and latency of AI computation. [3,4] In a crossbar array of analog memory components, doing an efficient multiplication operation requires large number of conductance states achievable almost linearly with change of voltage, current or time. In practical FE devices, achieving extremely high bit precision with significant linearity remained a challenge. By proper control of ferroelectric domain rotation [35], designing a custom-gate stack or by utilizing hybrid CMOS-FE synaptic cells, it is possible to achieve high number of intermediate conductance states, leading to more than 96% simulated accuracy [37] on classification tasks performed on MNIST handwritten dataset. In simulation, however, we made a simplified assumption that devices at each cross point is identical. For designing a full-hardware implementation, device to device variation, leakage and fatigue issues needs to be considered as well. While high FE polarization is essential for large device On/off ratio and opening a significant memory window, high leakage current results in significant static power loss increasing the power consumption and heating of the circuit. A leaky capacitor in a FeFET gate stack increases gate leakage current lowering the input impedance of the FeFET. Higher off-state current in an array could cause write or read disturbance to the nearby memory cells. For an 1T-1C FeRAM based synaptic circuit, there will be charge leakage that will lead to loss of data and need of frequent refreshing of the capacitor. Additionally, high gate leakage leads to earlier fatigue and breakdown of the FeFETs, that has catastrophic consequences of hardware failure. The current work suggests less leakage path through ferroelectric and less vacancy formation through low temperature annealing can lead to almost unlimited write endurance in BEOL ferroelectric analog memory and synaptic devices. A proper in-operando defect characterization and modelling is essential for a more accurate prediction of large-scale circuit implementation.

**Table 1.** Table benchmarking the present work with the state-of-the art results from HZO based ferroelectric capacitors focusing on the $P_r$, MW and endurance properties.

| Device architecture (BE/HZO/TE) | Annealing Temperature (°C) | Applied Electric field (MV/cm) | $P_r$ ($\mu C/cm^2$) | MW $2E_c$ (MV/cm) | Endurance (cycles) | CMOS BEOL compatible | Recovery ($2P_r$/ $2P_r$ Pristine) | Reference |
|---|---|---|---|---|---|---|---|---|
| W/HZO/W/Pt | 700 PMA 60 s | 4 @ 111 kHz (after $10^2$ cycles) | 32 | 4 | $10^8$ (111 kHz) | No | - | [11] |
| TiN/HZO/Al$_2$O$_3$/TiN | 400 PDA 30 s | 3.89 @ 1 kHz | 15.3 | 4 | - | Yes | - | [18] |
| TiN/HZO/Al$_2$O$_3$/Au | 500 PDA 30 s | 5 @ 500 Hz 3 @ 500 Hz | 30 12.5 | 6.2 - | $10^4$ (100 kHz) $10^6$ (100 kHz) | Yes | - | [21] |
| TiN/HZO/TiN | 600 PMA 30 s | 3 @ 1 kHz (after $10^5$ cycles) | 19.1 | 1.97 | $1.6*10^7$ (10 kHz) | No | - | [14] |
| TiN/ZrO$_2$-HfO$_2$ (superlattice)/TiN | 600 PMA | 3 (after $10^4$ cycles) | 15 | 3 | $10^{11}$ (1 MHz) | No | - | [23] |
| TiN/HZO/WN$_x$/Ru | 400 PMA 1 h | 2 @ 1 kHz | 12.5 | 3 | >$10^{10}$ (1.67 MHz) | Yes | 82% | [32] |
| TiN/HZO/Al$_2$O$_3$/Ti/Au | 450 PDA 30 s | 4 @ 1 kHz | 15 | < 3 V | >$10^9$ (100 kHz) | Yes | 88% | This work |
| TiN/TiNO$_X$/HZO/Al$_2$O$_3$/Ti/Au | 450 PDA 30 s | 5 @ 1 kHz | 21 | < 5 V | >$10^8$ (100 kHz) | Yes | 93% | This work |

## 4. Conclusions

In conclusion, microcrystalline structure and interface engineering for ferroelectric HZO capacitor devices is presented in this work. It is found that low thermal budget annealed samples possess high endurance performance exceeding $10^9$ cycles, however with a visible fatigue with increasing field cycling arising from oxygen vacancy redistribution related leakage currents. On minimizing the oxygen vacancy defects, devices grown and annealed with same parameters show improvements on $P_r$ and leakage currents, however, with earlier set in of fatigue. Both kinds of fatigue could be recovered, up to 90% for multiple times in our devices with a few pulse cycling that provides opportunity of runtime fatigue recovery in NVM components. Our results show that proper designing of stack and nanofabrication conditions can lead to CMOS BEOL compatible ferroelectric components with low fabrication cost, low operating cost and long worktime. Future experiments on optimized FE-DE stack, innovative device architectures and their large-scale integration will unravel the true potential of the technology.

## 5. Experimental Section/Methods

**Fabrication methods:** The Metal-FE-Insulator-Metal capacitors, with structures shown in **Figure 1(a)**, were fabricated using Atomic Layer Deposition (ALD). The bottom TiN electrode was grown by plasma-enhanced ALD (PEALD). HZO (10 nm) and Al$_2$O$_3$ (1.2 nm) layer was fabricated using thermal ALD. TiCl$_4$ and ammonia precursors were used for TiN growth, while HZO films were deposited with tetrakis (dimethylamino) hafnium and tetrakis (dimethylamino) zirconium as the Hf and Zr precursors, respectively, and water (H$_2$O) as the oxidant. 50:50 ratio of the Hf:Zr was obtained using alternate cycles of the Hf and Zr precursors. Following the HZO deposition, $Al_2O_3$ capping layer was deposited at 200 °C without breaking vacuum. After the film deposition, the sample was cleaved into one-quarter of a wafer and RTA was carried out for 30 seconds at annealing temperatures of 450 °C, 550 °C and 600 °C under nitrogen atmosphere (samples are named S1, S2 and S3 respectively). Finally, the FE capacitor structure was completed by evaporating Ti/Au (5/50 nm) as top electrode through electron beam evaporation. For sample S4, same fabrication condition as S1 was applied with the exception that TiN was left in air (in cleanroom ISO 4-5 cleanliness) for 7 days before HZO fabrication.

**Structural Characterization**: Structural characterizations of the samples were carried out by grazing incidence X-ray diffraction technique (GIXRD) in Rigaku SmartLab diffractometer operating at 45 kV, 150 mA, with a Cu rotating anode. The incident angle ω was set at where the angle was at half maximum reflectivity intensity. ALD technique ensures the atomically flat surface. For our samples, this is confirmed by atomic force microscopy (AFM) with the root mean square roughness of around 0.5 nm.

**Electrical Characterization**: Electrical characterizations of the samples were carried out by applying voltage between the bottom TiN and the top gold electrode. For ferroelectric polarization – electric field (*P–E*) hysteresis, endurance and retention measurements, ferroelectric material tester AixACCT 2000E was used. Static current – voltage (*I-V*) characteristics were measured with a semiconductor parameter analyzer Keysight 1500A. Ferroelectric hysteresis was measured on 200 x 200 µm² square junctions using both dynamic hysteresis measurement (DHM) and Positive-up-negative-down (PUND) techniques with 1 kHz of triangle voltage sweep. For endurance measurement, a rectangular voltage pulsing of 3 V, 100 kHz frequency was applied followed by 3 DHM measurements per decade to monitor the $P_r$ value of the sample. All measurements were carried out at room temperature and under ambient conditions.

**Acknowledgements**
The authors acknowledge financial support from Research Council of Finland through projects IntelliSense (no. 345068), AI4AI (no. 350667), Ferrari (no. 359047), Business Finland and European Commission through project ARCTIC (no. 101139908). The work used experimental facilities of OtaNano Micronova cleanroom and Research Council of Finland Infrastructure "Printed Intelligence Infrastructure" (PII-FIRI, project no. 358618).

Supporting Information

**Designing high endurance ferroelectric memory and neuromorphic hardware through engineered recovery of fatigue in $Hf_{0.5}Zr_{0.5}O_2$ thin films**

Xinye Li, Padma Srivari and Sayani Majumdar *

Faculty of Information Technology and Communication Sciences, Tampere University, 33720 Tampere, Finland

E-mail: sayani.majumdar@tuni.fi

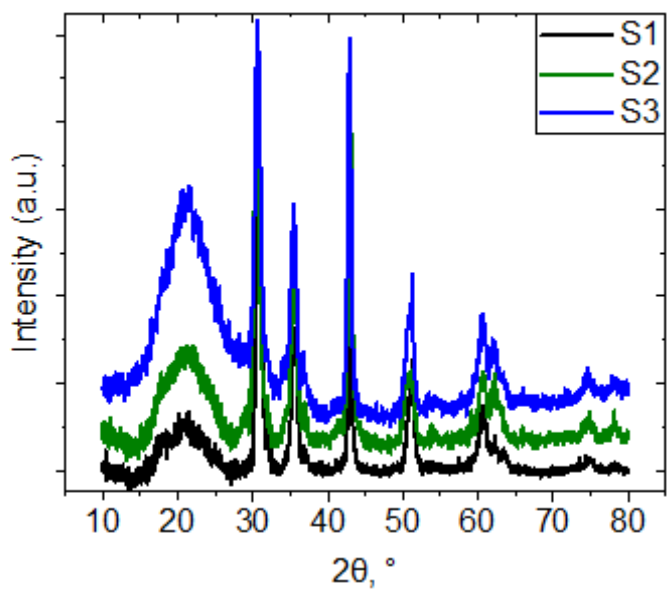

**Figure S1.** Long range scan of 2θ from 10° to 80° by GIXRD for samples S1, S2 and S3.

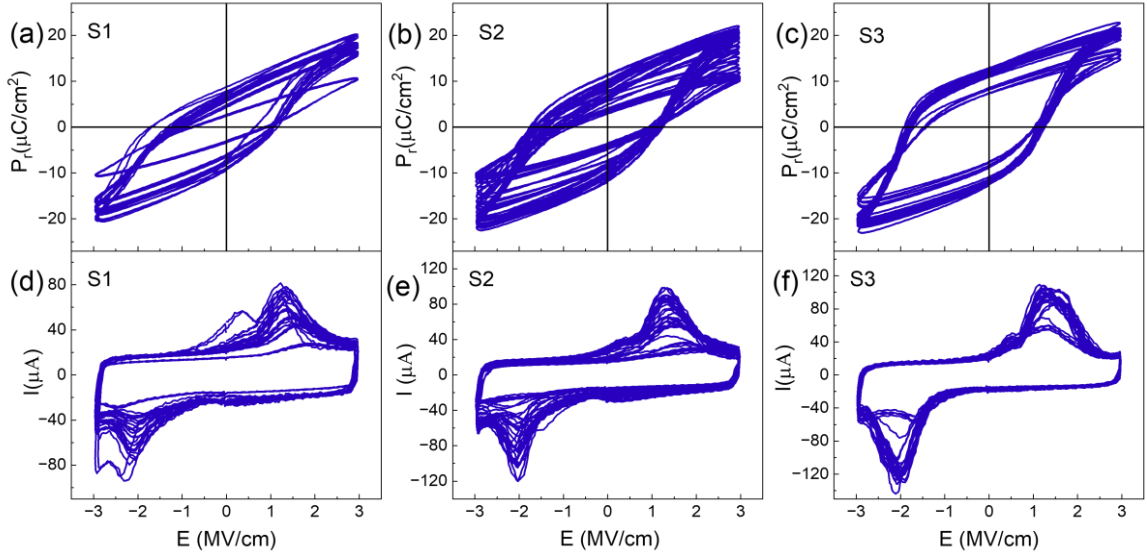

**Figure S2.** DHMs measured from devices at multiple different locations of the wafer from S1, S2 and S3 for a quick comparison.

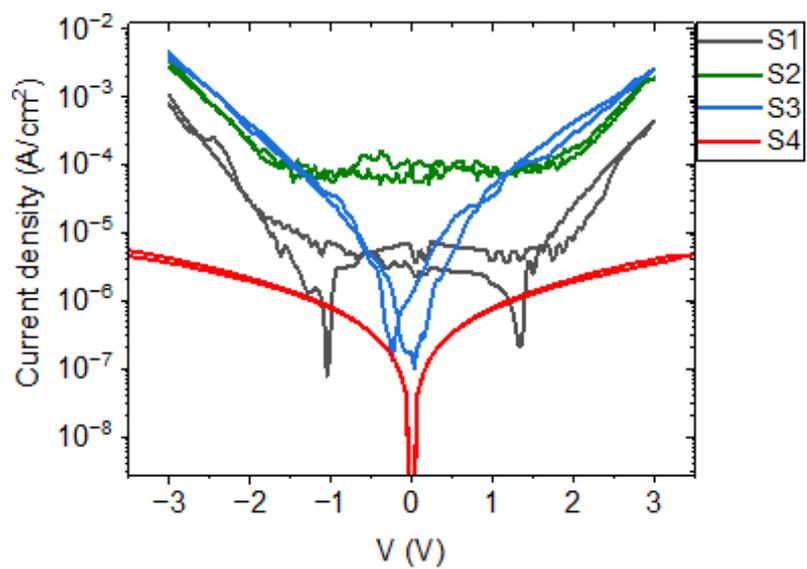

**Figure S3.** Static Current density -Voltage characteristics of samples S1, S2, S3 and S4.

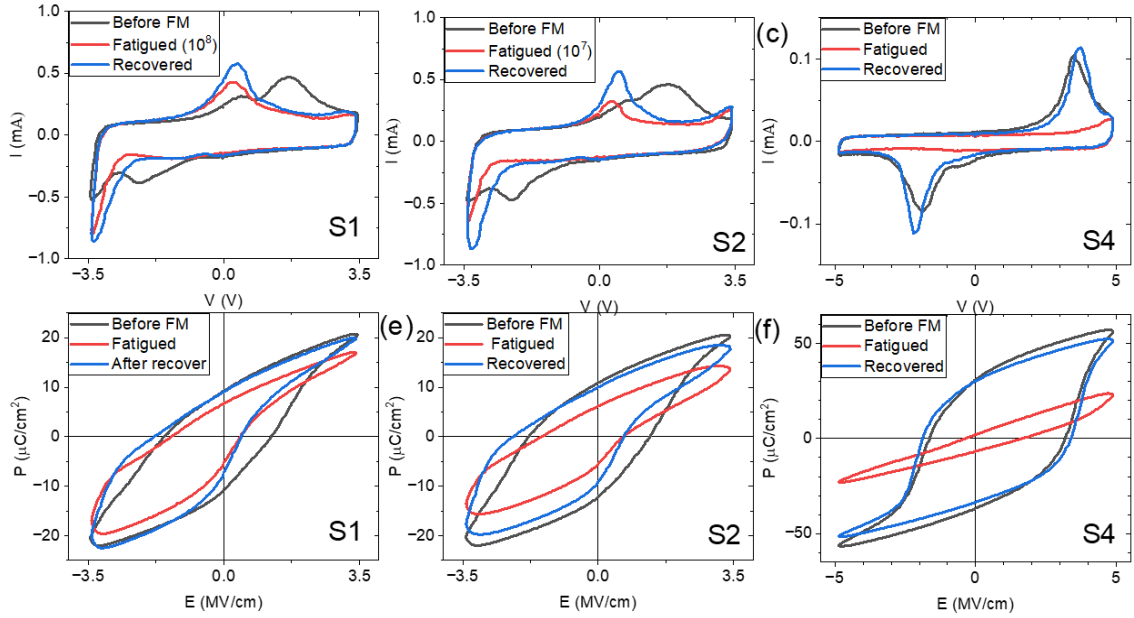

**Figure S4.** DHMs measured before fatigue pulsing cycle (black lines), after certain number of fatigue cycles (red lines), after recovery DHMs (blue lines) of S1, S2 and S4. (**a,b,c**) are I-V plots; (**d,e,f**) are P-E plots.

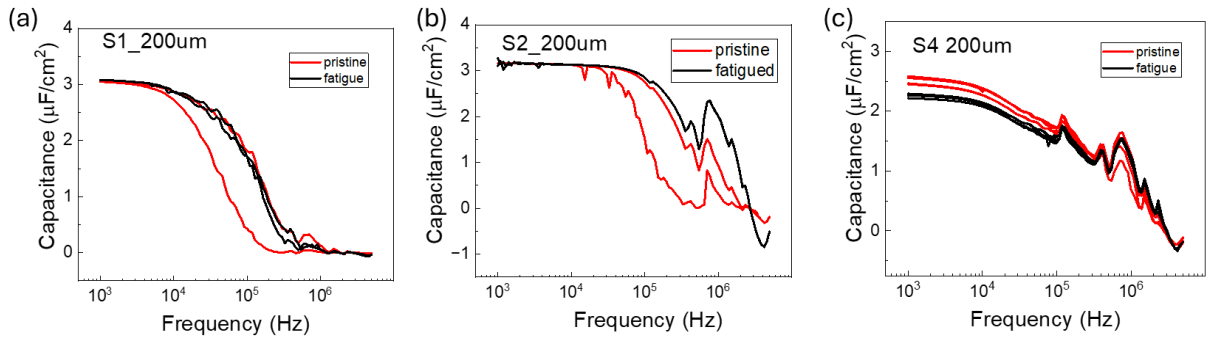

**Figure S5.** Capacitance (*C*) per unit area as a function of frequency (*f*) for sample S1 (a), S2 (b) and S4 (c) in both pristine and fatigued states.